\documentstyle[12pt]{article}

\newcommand{\eq}{\begin{equation}}
\newcommand{\en}{\end{equation}}
\newcommand{\eqa}{\begin{eqnarray}}
\newcommand{\ena}{\end{eqnarray}}

\newcommand{\NP}[1]{Nucl.\ Phys.\ {\bf #1}}

\newcommand{\PR}[1]{Phys.\ Rev.\ {\bf #1}}

\begin{document}

\hskip 11.5cm \vbox{\hbox{DFTT 71/92}\hbox{December 1992}}
\vskip 0.4cm
\centerline{\bf   THE KAZAKOV-MIGDAL MODEL}
\centerline{\bf   AS A HIGH TEMPERATURE LATTICE GAUGE THEORY}
\vskip 1.3cm
\centerline{ M. Caselle, A. D'Adda  and  S. Panzeri}
\vskip .6cm
\centerline{\sl Istituto Nazionale di Fisica Nucleare,Sezione di Torino}
\centerline{\sl  Dipartimento di Fisica
Teorica dell'Universit\`a di Torino}
\centerline{\sl via P.Giuria 1, I-10125 Turin,Italy}
\vskip 2.5cm

\begin{abstract}
We show that the Kazakov-Migdal (K-M) induced gauge model in $d$
dimensions describes the high temperature limit of ordinary lattice
 gauge theories in $d+1$ dimensions. The matter fields
are related to the Polyakov loops, while  the spatial gauge variables
become the gauge fields of the K-M model.
This interpretation of the K-M model is in agreement with some recent
results in high temperature lattice QCD.
\end{abstract}
\vskip 5cm
\hrule
\vskip1.2cm
\noindent

\hbox{\vbox{\hbox{$^{\diamond}${\it email address:}}\hbox{}}
 \vbox{\hbox{ Decnet=(31890::CASELLE,DADDA,PANZERI)}
\hbox{ internet=CASELLE(DADDA)(PANZERI)@TORINO.INFN.IT}}}
\vfill
\eject

\newpage

{\bf 1. Introduction}
\vskip 0.3cm

Recently V.Kazakov and A.Migdal proposed a new lattice gauge model, in
which the gauge self-interaction is induced by scalar fields in the
adjoint representation~\cite{KM}.
The action they propose is
defined on a generic  d-dimensional lattice
and has the following form:

\eq
S = \sum_{x} N {\rm Tr} \bigl[
 m^{2} \phi^{2}(x) - \sum_{\mu} \phi(x)U(x,x+\mu)\phi(x+\mu)
U^{\dagger}(x,x+\mu)\bigr]
\label{km}
\en
where $\phi(x)$ is an Hermitian $N \times N$ matrix defined on the sites
$x$ of the lattice
and
 $U(x,x+\mu)$ is a unitary $N \times N$ matrix, defined  on the links
$(x,x+\mu)$, and plays the role, as in the usual lattice
discretization of Yang-Mills theories, of the gauge field.
Integrating over the
scalar field $\phi$ one can induce an effective action for the gauge
field,
\eq
\int DU D\Phi exp (-S) \sim \int DU exp ( -S_{ind}[U])
\en
with:
\eq
S_{ind}[U] = - \frac{1}{2} \sum_{\Gamma} \frac{ |{\rm Tr} U[\Gamma]|^{2}}
{l[\Gamma] (2m^2)^{l[\Gamma]}}~~~,
\label{gauge}
\en
where $l[\Gamma]$ is the length of the loop $\Gamma$, $U[\Gamma]$ is
the ordered product of link matrices along $\Gamma$ and the summation
is over all closed loops.

The model turns out to be
solvable in the large $N$ limit, for any space-time dimension
$d$~\cite{KM,M,G} and
exactly solvable, for any value of $N$ in d=1~\cite{CAP}.
A further interesting feature of the model is its deep connection with
 the theory of non-critical strings, in particular it can be shown
 that in the $d=1$ case the K-M model is equivalent to the vortex free
 sector of the d=1 string~\cite{GK}.
Despite all these nice features the model seems to miss
the original goal of Kazakov and Migdal, since the induced gauge
theory~(\ref{gauge}), due to the fact that the matter fields are in the
adjoint representation,  has a super-confining behaviour~\cite{KSW} and
seems to have the wrong perturbative vacuum~\cite{KMSW,KMSW2}.
Several improvement of the original K-M action have been proposed to
avoid these problems~\cite{KMSW2,Mig92d,Kh-Ma,Ru}, but even if many
interesting results
have been obtained the problem of the identification with a pure
lattice gauge theory of the ordinary type (namely with the ordinary
confinement behaviour) while keeping exact solvability is still open.

In this letter we suggest a new point of view on the problem and show
that the K-M model can be identified {\it before integrating over the
matter fields} with an ordinary $SU(N)$ lattice
gauge theory in the high temperature limit, the matter fields being
related to the Polyakov loops, while the spatial gauge variables become
the gauge fields of the K-M model.

This letter is organized as follows: after a short introduction (just to
set notations)
on finite temperature lattice gauge theory (sect.2), we describe and
discuss the
equivalence with the K-M action  in sect.3 and 4.
Sect.5 is devoted to the continuum limit and sect.6 to
some concluding remarks.

\vskip 2cm
{\bf 2. Finite temperature lattice gauge theories}
\vskip 0.3cm

Let us consider a pure gauge theory with gauge group $SU(N)$ defined
on a $d+1$ dimensional cubic lattice. In order to describe a finite
temperature LGT we must take periodic boundary conditions in one
direction (which we shall call from now one ``time-like'' direction),
while the boundary conditions in the other $d$ direction (which we shall
call ``space-like'') can be chosen freely. Let us take a lattice of
 $N_t$
($N_s$) spacings in the time (space) direction.
The theory will contain only gauge fields described by the link
variables $U_{ n;i} \in SU(N)$ where $ n \equiv (\vec x,t)$ denotes the
 space-time position of the link and $i$ its direction.
It is useful to choose
different couplings in the time and space directions. Let us call them
$\beta_t$ and $\beta_s$ respectively. Let us take the simplest choice
for the lattice gauge action, namely the Wilson action:

\eq
S_W=\sum_{n}~N~Re\left\{\beta_t\sum_i~tr(U_{n;0i})
+\beta_s\sum_{i<j}~tr(U_{n;ij})\right\}
\label{wilson}
\en

where $U_{n;0i}$ ($U_{n;ij}$) are the time-like (space-like)
plaquette variables, defined as usual:

\eq
U_{n;ij}=U_{ n;i}U_{n+ \hat i;j}
U^\dagger_{n+\hat j;i}U^\dagger_{n;j}
\en

$\beta_s$ and $\beta_t$ are related to the  (bare) coupling constant $g$
 and the
temperature $T$ of the gauge theory by the relations \footnote { Notice
that the coupling constant $g$ is rescaled here by a factor $N$ with
respect for instance to the notations of ref.~\cite{sy} }:

\eq
\frac{2}{g^2}=a^{3-d}\sqrt{\beta_s\beta_t},
\hskip 1cm
T=\frac{1}{N_ta}\sqrt{\frac{\beta_t}{\beta_s}}
\en
where $a$ is the space-like lattice spacing while $\frac{1}{N_tT}$ is
 the time-like spacing. The two are related by the adimensional ratio
$\epsilon\equiv \frac{1}{N_tTa}$.
We can solve the above equations in terms of $\epsilon$ as follows:
\eq
\beta_t=\frac{2}{g^2\epsilon}a^{d-3}
\label{betat}
\en
\eq
\beta_s=\frac{2\epsilon}{g^2}a^{d-3}~~.
\en

In a finite temperature discretization it is possible to define
gauge invariant variables which are topologically non-trivial loops,
closed due to the periodic boundary conditions in the time directions.
The simplest choice is the Polyakov loop defined as follows:

\eq
P(\vec x)= tr \prod_{t=1}^{N_t}(U_{\vec x,t;0})
\label{polya}
\en

where x labels the space coordinates  of the lattice sites.
Moreover, an important feature of the finite temperature theory with
respect to the zero temperature case is that it has a new global
symmetry (independent from the gauge symmetry) with symmetry group the
center $C$ of the gauge group (in our case $Z_N$). The Polyakov loop
turns out to be a natural order parameter for this symmetry.

It is possible to obtain, just from the definition itself of the model,
some general properties in the high temperature regime (see for
instance~\cite{sy}). In this region the symmetry with respect of the
center of the group is broken, the theory is deconfined, the Polyakov
loop has a non-zero expectation value and, what is more important, it is
an element of the center of the gauge group.
Physically this means that the
 links in the time direction fluctuate
around one of the $N^{th}$ roots of unity of the $Z_N$ group. We can
 lift this degeneracy by adding to the Wilson action a ``magnetic term''
\eq
S_m= N~ h~\sum_{\vec x} Re ~tr \prod_{t=1}^{N_t}(U_{\vec x ,t;0})
\label{sm}
\en
and eventually send $h\to 0$.
This procedure selects the vacuum in which

\noindent
$<P(\vec x)>={1}$.

The spatial links are
static up to gauge transformations and this fact tells us that the
spatial degrees of freedom of the model behave as if they belonged to a
zero temperature model in one dimension less and with coupling constant
$g^2T$. If one could integrate out these spatial degrees of freedom the
resulting effective action for the Polyakov loops would be that of a d-
dimensional spin model with short range spin-spin interaction~\cite{sy}.

\vskip 2cm
{\bf 3. The K-M model as an high temperature LGT.}
\vskip 0.3cm

Let us take the $d+1$ dimensional LGT described above (with gauge
group $SU(N)$) and consider first a very special case,
namely $N_t=1$. We will be
interested in the
 $T\to \infty$ (hence $\epsilon\to 0$) limit.

Let us call $S_t$ ($S_s$) the contribution of the temporal (spatial)
 plaquettes
to the action, with a ``magnetic term'' of the type (\ref{sm}) included
in $S_t$.
Due to the boundary conditions and to the $N_t=1$ position we can
rewrite $S_t$ as:

\eq
S_t=N~\beta_t \sum_{\vec x,i} ~Re~
tr\left\{U_{\vec x;i}U_{\vec x+ \hat i;0}
U^\dagger_{\vec x;i}U^\dagger_{\vec x;0}\right\}
+ N~  h~\sum_{\vec x} Re~ (U_{\vec x;0})
\label{ht}
\en

This model is interesting in itself and represents the generalization of
the K-M model to ``matter'' fields described by unitary matrices.
Notice also that since  the
space contribution $S_s$ to the action  is depressed as $\beta_s\sim 1
/T$, in the $T\to \infty$ limit this model should capture most of the
features of  high temperature LGT.

Following Svetitsky and Yaffe ~\cite{sy} one can argue at this point
that in the high temperature limit each temporal plaquette fluctuates
around the identity . It is easy to see that the fluctuations are of
 order $1/\sqrt{\beta_t}$ . In fact if we put
$U_{pl} = e^{i \frac{\phi}{\sqrt{\beta_t}}}$
with $\phi$ a traceless hermitian matrix
we have:
\eq
\int dU_{pl} e^{N\beta_t tr~Re~U_{pl}} = \frac{e^{N^2
\beta_t}}{\sqrt{\beta_t}} \int d\phi e^{-N~ tr \phi^2} [ 1+ O(1/\beta_t)
]
\en
\label{upl}
Correspondingly the link variables in the time direction fluctuate
around one of the $N^{th}$ roots of unity of the $Z_N$ group with
fluctuations still of order $1/ \sqrt{\beta_t}$. When the
magnetic term is switched on the vacuum is forced to be
 $U_{\vec x;0}={\bf 1}$ and we can expand $U_{\vec x;0}$ in the
 following way:

\eq
U_{\vec x;0}\equiv e^{i\frac{\phi(\vec x)}{\sqrt{\beta_t}}}
={\bf1}+i\frac{\phi(\vec x)}{\sqrt{\beta_t}}
-\frac{\phi^2(\vec x)}{2\beta_t}
+\cdots
\label{svil}
\en
 By inserting (\ref{svil}) in (\ref{ht}) we find:

\eq
S_t=N~\beta_t~~tr\left\{d~\Omega_{x}
{\bf 1}+\frac{1}{ \beta_t}\sum_{\vec x} \left(-m^2 \phi(\vec x)^2+
\sum_{i=1}^d U_{\vec x;i}\phi(\vec x)U^\dagger_{\vec x;i}
\phi(\vec x+\hat i)\right)\right\}
\label{htkm}
\en
where $m^2=d+\frac{h}{2 \beta_t}$ and $\Omega_x$ is the volume of the
$d$ dimensional space.
In ref.~\cite{sy} only the constant leading contribution of order
$\beta_t$ was considered; here we recognize that the contributions of
the fluctuations, of order $1$, coincide, apart from an irrelevant
 constant, with the K-M action with arbitrary $m^2$
(eq. (\ref{km})) and in the limit of zero magnetic field with the K-M
action at the critical point $m^2=d$. The expansion (\ref{svil}) and in
particular the choice of the expansion parameter
 $\frac{1}{\sqrt{\beta_t}}$ is also justified {\it a posteriori} by
 the
analysis of the distribution of the eigenvalues of $\phi(\vec x)$
 resulting
from the action (\ref{htkm}). It was shown in ~\cite{G} that at least in
the large $N$ limit and for $m^2>d$ the eigenvalue distribution of $\phi
(\vec x)$ is semicircular with a finite , non-zero radius.
 This implies that the eigenvalues of the original unitary matrix
 $U_{\vec x;0}$ are restricted
to a region of order $\frac{1}{\sqrt{\beta_t}}$ around $1$ on the unit
 circle, thus justifying the expansion (\ref{svil}).
For $d>1$ such distribution of the eigenvalues survives also in the
limit $h \to 0$ (with $h>0$) namely $m^2 \to d$, whereas for $d=1$ the
radius of the eigenvalue distribution goes to infinity as $h \to 0$
($m^2 \to 1$) .
This shows that in the one dimensional case the critical point can only
be described in the unitary matrix formulation given by (\ref{ht}).

Let us consider now the case of an arbitrary $N_t$. The action (without
the contribution of spatial plaquettes) is:
\eq
S_t = N\beta_t \sum_{\vec x,i} \sum_{t=1}^{N_t} Re~tr\{ U_{\vec x,t;i}
U_{\vec x +\hat i,t;0}U^{\dagger}_{\vec x,t+1;i}U^
{\dagger}_{\vec x,t;0} \} +N~h~\sum_{\vec x}Re P(\vec x)
\label{nt}
\en

where $P(\vec x)$ is the Polyakov loop given in (\ref{polya}).
It is possible to choose the gauge in such a way that $U_{\vec x,t;0}
=1$ for $t=1,2,..,N_t -1$ so that the only non trivial links in the
time direction are $U_{\vec x,N_t;0}$ , related to the Polyakov loop by
the relation $P(\vec x) = tr U_{\vec x,N_t ;0}$.
The plaquettes at $t<N_t$ then reduce to $U_{\vec x,t;i}U^
{\dagger}_{\vec x ,t+1;i}$ . It follows that in the large
$\beta_t$ limit the spatial link variables at different values of $t$
coincides up to fluctuations of order $1/\sqrt{\beta_t}$. So we can put
\eq
U_{\vec x,t;i} = U_{\vec x;i} exp\{i\frac {\psi_i (\vec x,t)}{\sqrt
{\beta_t}}\}
\label{usp}
\en
where we are free to choose $\psi_i(\vec x,1)=0$.
In complete analogy with the case $N_t=1$ we also have:
\eq
U_{\vec x,N_t;0} = exp\{i \phi (\vec x) \sqrt{\frac{N_t}{\beta_t}}\}
\label{utemp}
\en
where the factor $\sqrt{N_t}$ in the exponent at the r.h.s. of
 (\ref{utemp}) is needed , as we shall see, for a smooth $N_t \to \infty
$ limit.
By inserting equations (\ref{usp}) and (\ref{utemp}) into the
action (\ref{nt}) and by expanding  in powers of $1/\beta_t$ up to terms
of order $1$ , one obtains an action quadratic in the
$\psi_i(\vec x,y)$'s:
\begin{eqnarray}
S_t&=&N~ tr \sum_{\vec x;i}\{N_t \beta_t +\sum_{t=2}^{N_t-1}
 [\psi_i(\vec x,t)\psi_i(\vec x,t+1) - \psi^2_i(\vec x,t)] -
 \psi^2_i(\vec x,N_t) -\nonumber \\&& -\sqrt{N_t}\psi_i(\vec x,N_t)
[\phi(\vec x+\hat i)-U^{\dagger}_{\vec x;i}\phi(\vec x)U_{\vec x;i}]
 -\frac{N_t}{2} [\phi(\vec x+\hat i)- \nonumber \\&&-
U^{\dagger}_{\vec x;i}\phi(\vec x)U_{\vec x;i}]^2  -Nh\frac{N_t}{\beta_t}
\phi^2(\vec x)\}
\label{uff}
\end{eqnarray}
The fluctuations $\psi_i(\vec x,t)$ can be eliminated by performing the
corresponding gaussian integrals and the final result coincides with the
K-M model as given by eq.(\ref{htkm})\footnote {Notice that $\Omega_x$
in (\ref{htkm}) is now replaced by the volume of the whole space-time
lattice}.
The same result can be obtained also by noticing that, as the self
interaction term is neglected, each spatial link variable
 $U_{\vec x,t;i}$ belongs to just two temporal plaquettes and that the
corresponding integral can be done by using character expansion in
analogy to lattice QCD in two dimensions. This allows to define a
``renormalized'' coupling $\beta_{eff}$ through the relation:
\eq
I_f(\beta_{eff})=\left[\frac{I_f(\beta)}{I_0(\beta)}\right]^{N_t}
\en
where $I_f$ and $I_0$ are the coefficients of the  fundamental
and identity characters in the character expansion of the action.
Then it can be shown that the large $\beta$ behaviour of
$\beta_{eff}(\beta)$ is as expected $\beta_{eff}\sim \frac{\beta}{N_t}$
(see for instance ref.~\cite{dz}).

The final result eq.(\ref{htkm}) is then independent from the number
$N_t$ of lattice spacing in the time direction provided the field $\phi(
\vec x)$ is normalized as in (\ref{utemp}). Notice that the r.h.s. of (
\ref{utemp}) is also independent of $N_t$ according to the definition of
$\beta_t$ given in (\ref{betat}). So the continuum limit in the time
dimension $N_t \to \infty$ is well defined and leads to the K-M model.

Hence we can conclude that
\vskip 0.3 cm
{\sl for $d>1$ the $d$-dimensional K-M model is a good description of the
 small fluctuations of the Polyakov loops around their minimum (frozen)
 position in the high temperature limit of a pure lattice gauge theory
in $d+1$ dimensions with the Wilson action.}

\vskip 2.0 cm

{\bf 4. Comments and remarks}

\vskip 0.2 cm

The integration over the $\phi$ fields in (\ref{htkm}) leads to the
induced gauge action eq.(\ref{gauge}) which can now be interpreted as
describing the spatial degrees of
freedom of a pure LGT  at $T=\infty$. Since,
as mentioned above, the space degrees of freedom in the
high temperature limit behave as a zero temperature d-dimensional gauge
theory with coupling constant $g_{eff}^2=g^2T$, the K-M model
corresponds to the $g_{eff}=\infty$ point of the theory thus leading to
the superconfinement behaviour.

At $T<\infty$ one has to add a small (order $1/T$) gauge
 self-interaction term.
This term obviously destroys the exact solvability of the model,
but it can be treated as a small perturbation around the K-M solution.
The main outcome of our analysis is probably in the fact that we now
understand that
such a perturbative analysis would be an high-temperature expansion for
the LGT.

Let us consider for instance the expectation value of a non backtracking
Wilson loop which is zero at $T=\infty$ due to the
superconfinement. Its first non trivial perturbative
contribution in the $1/T$ expansion is obtained  by filling the loop by
elementary plaquettes \footnote{In more than two spatial dimension
one has to sum  over the contributions of all surfaces made of
 elementary plaquettes and having the
original loop as a boundary. These contributions are of order $T^{- \cal
A}$ where $\cal A$ is the area of the surface, so that the surface of
minimal area dominates at high temperature.}. The resulting
 ``filled Wilson loops'' is invariant under the local $Z_N$ symmetry
 and has to be evaluated using the measure of the pure M-K induced
 action. This was done in ref.~\cite{KMSW} where it was shown that the
``filled Wilson loop" has an area law behaviour.
It is also remarkable and not quite understood yet that in this context
 intriguing connections with 2d spin models emerge. Our interpretation
is that the ``filled Wilson loop " is the first non
vanishing term in a large temperature expansion of the vacuum
expectation value of the ordinary Wilson loop.

The K-M model is {\it only}  apt to describe  fluctuations around
one given vacuum of the high temperature LGT; in other words it can only
describe finite temperature LGT in the broken $Z_N$ phase.
However, even in the framework of the K-M model one has at least a
qualitative understanding of the restoring of the $Z_N$ symmetry below
the critical temperature.
In fact the expectation value of the Polyakov loop
\eq
<P(\vec x)> = <Tr e^{i\phi(\vec x)\sqrt{\frac{N_t}{\beta_t}}}>
\en
can be evaluated in the large $N$ limit by replacing $\phi(\vec x)$ with
its classical value given in ref.~\cite{G}. At the critical point $m^2=
d$ this is given by a semicircular distribution of eigenvalues of radius
$r=\sqrt{\frac{2d-1}{d(d-1)}}$. At high temperature, more precisely for
$r\sqrt{\frac{N_t}{\beta_t}}\ll \pi$, the eigenvalues are
 peaked around one vacuum and the $Z_N$ symmetry is broken.
 However for values of $\beta_t$ such that
$r\sqrt{\frac{N_t}{\beta_t}}\approx \pi$ the eigenvalues
 of $\sqrt{\frac{N_t}{\beta_t}}\phi(\vec x)$ are distributed over
several vacua on most of the unit circle  and so the symmetry is
 eventually restored.

Without a more detailed understanding of the symmetry restoration
 nothing new can be said in this context on the Svetitsky-Yaffe
 conjecture that the
critical behavior of a $(d+1)$-dimensional finite temperature gauge
theory is the same as that of a $d$-dimensional spin model with the same
$Z_N$ symmetry. At this stage at least the  emergence of spin models
 ~\cite{KMSW} in the expectation value of the Wilson loop does not
 appear to be correlated with the Svetitski-Yaffe conjecture.
 It should be
noticed however that if the contribution of the
 spatial plaquettes is neglected, an effective theory for the Polyakov
 loop valid also in the region of large $\phi(\vec x)$
could in principle be obtained directly from (\ref{nt}) by integrating
 over the spatial links variables.
 This would require the use of a generalization of the Itzykson-Zuber
formula described in ~\cite{KMSW} which is exact for $N=2$ and valid  as
an asymptotic formula for $N>2$.

The instability which occurs in the $m^2<d$ region, corresponding to
negative values of $h$ in (\ref{nt}), can be understood
 from the high temperature LGT point of view as a consequence of the
 fact that  with our ``magnetic field'' in this regime we are actually
 pushing the system $out$ of the chosen vacuum which becomes unstable.
Similarly the instability due to the presence of a linear
term~\cite{lat92} which occurs if one looks at U(N) instead of SU(N) K-M
models, can be understood as a signature of the Goldstone modes of the
broken U(1) symmetry which one has in this case.

Let us finally consider the case $d=1$. Spatial plaquettes are absent in
this case and eq. (\ref{nt}) gives the complete action irrespective of
the value of $\beta_t$. This is just two dimensional QCD on a cylinder
or , in case the spatial dimension is also compactified, on a torus.
 We know from ref.~\cite{G} that in the $d=1$ case at the critical
 point $m^2=1$ the distribution of the eigenvalues of $U_{\vec n,N_t;0}$
 is not confined to a small region of the unit circle. As a consequence
we are always in the unbroken phase and the description of the
system in terms of the K-M model is not valid.
On the other hand by expanding  both spatial and temporal links in
 a fashion similar to eq.(\ref{usp}) one can reduce the action of eq.
(\ref{nt}) with periodic boundary conditions in both space and time
 to an action
containing only one space and one time-like link, in agreement with
general result on lattice QCD in two dimensions~\cite{Ru}.

\vskip 2cm
{\bf 5. Taking the continuum limit}
\vskip 0.3cm

In the previous sections we have derived the K-M model as a high
temperature limit of lattice QCD.
However while in ordinary lattice QCD one has a well defined
continuum limit, this is not the situation in
the K-M model with a quadratic potential at the critical point $m^2=d$.
It was pointed out in ~\cite{G} that if one approaches $m^2=d$ from
below there exists a solution of the master field equation corresponding
to a semicircular distribution of eigenvalues whose radius $r$ goes to
infinity as $m^2 \to d$, a signal of the existence of a critical point.
Such configuration however is a local maximum of the free energy;
moreover the free energy itself is unbounded from below for $m^2<d$.
This can be cured by adding a higher order term to the action, for
instance a quartic term $\frac{\lambda}{4} Tr\phi^4$.
{}From the point of view of the high temperature expansion this is not
such an {\it ad hoc} adjustment as it might appear at first, in fact the
continuum limit corresponds to the limit where the radius of the
eigenvalue distribution goes to infinity and in that limit higher order
terms in $\phi(\vec x)$ are expected to be relevant. Therefore in the
continuum limit the high temperature QCD is most likely described by
a K-M model with a higher order potential in agreement, as discussed
below, with some recent results on lattice QCD~\cite{ht1,ht2}.
 It has indeed been shown in~\cite{ags} that, with the addition of a
 quartic term  $\frac{\lambda}{4} Tr\phi^4$ to the action, a
second order critical point can be reached, at least in the case of
 SU(2) in d=4.
 The analysis of the phase
diagram can be done rather easily within a mean field approximation:
in the $\lambda,m^2$
plane one  finds a line of first order transitions starting from the
K-M critical point and ending with a second order critical point
located at $\lambda=2.57$ and $m^2=2.26$\footnote{Notice that there is a
factor of two between our normalization of the mass parameter and that
of~\cite{ags}}. Moreover direct Montecarlo
simulations show that these mean field results are quite accurate and
even the critical point location is essentially confirmed by the
computer simulation. We have made a similar mean field analysis in the
$d=3$ case, in which we are interested, showing a similar scenario: a
line of first order phase transitions starting from $m^2=3$ and ending
with a
second order critical point located at $\lambda\sim2.2, m^2\sim 1.56$.
In~\cite{ags} it was stressed that the corresponding continuum theory
had nothing to do with ordinary SU(2) gauge theory. Our suggestion is
that {\sl it should instead describe the high temperature, deconfined
 phase of SU(2)} (with one more space-time dimension). This conjecture
 is in
remarkable agreement with some recent results on the high temperature
behaviour of lattice gauge theories, both within a perturbative
approach~\cite{ht1}, and with montecarlo simulations~\cite{ht2}.
The main idea behind~\cite{ht1,ht2} is that (3+1) dimensional gauge
theories at high temperature undergo a peculiar form of dimensional
reduction. The original picture~\cite{ht0} of a complete dimensional
reduction (namely a complete decoupling of the degrees of freedom in the
compactified time direction, which would lead to an effective
three-dimensional theory of the ordinary type) does not occur in
 general and one finds instead a three dimensional gauge theory coupled
 with a scalar
field $\phi$ in the adjoint representation with a non trivial potential
$V(\phi)$. This potential has a quadratic term with the wrong sign to be
identified as a mass term and a quartic contribution which stabilizes
the action, exactly as in our mean field solution. Moreover the action
used in~\cite{ht2} (eq.(20) of~\cite{ht2}) for the montecarlo simulation
is exactly that of a K-M model in the instable ($m^2<3$) region, with a
quartic term (like the above described mean field solution) plus the
gauge self-interaction term. Taking into account the obvious
uncertainties of our mean field estimate the numerical agreement on the
coefficient of the quartic contribution between our result and that
of~\cite{ht2} is impressive.

\vskip 2cm
{\bf 6. Conclusions}
\vskip 0.3cm

In this letter we have shown that, besides the usual identification with
a LGT coupled with adjoint matter  in the strong coupling regime,
the Kazakov Migdal model can also be related to the behaviour of a pure
lattice gauge theory (with Wilson action) in the limit of very high
temperature. In particular we have seen that the $\phi$ fields describe
the small fluctuations around the frozen position of the Polyakov loops.

We think that this new point of view is  important if we want to
understand some of the peculiar features of the K-M model
(superconfinement, lack of
ordinary perturbative vacuum, presence of an unstable phase for $m^2<d$
and of a finite gap in the free energy), but besides this we think that
there are two other reasons of interest.
First the master field solution of the K-M model could  give us a
powerful tool to do both analytical and numerical calculations in
 pure LGT at high temperature.
Second, more ambitious point: the K-M model is related to non
critical strings,  although this relation is still not quite clear for
 $d>1$ (see however ~\cite{bou} and ~\cite{dakl} for some recent
progress on this subject). Our interpretation
 might lead (at least in the high temperature, large N limit) to a first
 description in terms of strings of LGT, a hope which
justifies by itself further efforts toward a better understanding of the
K-M model.

\vskip 2cm
{\bf Note added}
\vskip 0.3cm
While completing this letter we received  a new interesting paper by
Dobroliubov, Kogan, Semenoff and Weiss~\cite{dksw}, in which the
interpretation of the ``filled Wilson loop'' as the result of the
perturbative expansion of gauge self-interaction term is discussed.
 The two regimes discussed in~\cite{dksw} may be identified
within our approach as the high temperature deconfined regime and the
low temperature confined phase respectively, and the coupling $\lambda$
of~\cite{dksw} with our ratio $\frac{\beta_s}{\beta_t}$
\footnote{Notice, in order to
avoid confusion that our high temperature phase is what is called
in~\cite{dksw} the low temperature one and is characterized by a K-M
behaviour and by $\lambda\equiv\frac{\beta_s}{\beta_t}\to 0$}.

\vskip 2cm
{\bf Acknowledgments}
\vskip 0.3cm
One of us (A.D.) thanks J. Ambj{\o}rn and D. Boulatov for useful
discussion and in particular J. Ambj{\o}rn for pointing out to us the
relevance of ref. ~\cite{ht1} and ~\cite{ht2}.

\end{document}